\documentclass[aps,prl,twocolumn,groupedaddress,showpacs]{revtex4}
\usepackage{graphics}
\usepackage{amsfonts}
\usepackage{amsmath}
\usepackage{epsfig}
\usepackage{epsf}
\usepackage{graphicx}
\usepackage{dcolumn}
\usepackage{bm}
\usepackage{color}
\usepackage{array}

\begin{document}

\title{Two-photon exchange corrections to $\gamma^*N\Delta$ form factors for  $Q^2\leq 4 $ (Gev/$c$)$^2$ }
%\title{Two-Photon Exchange and Pion Electromagnetic Production at $\Delta(1232)$ Peak}
\author{
Hai-Qing Zhou$^1$ and Shin Nan Yang$^2$ \\
$^1$Department of Physics,
Southeast University, NanJing 211189, China\\
$^2$Department of Physics and Center for Theoretical Sciences,\\
National Taiwan University, Taipei 10617, Taiwan}

\date{\today}

\begin{abstract}
We evaluate the corrections of the two-photon exchange (TPE) process on  the $\gamma^* N\Delta$ transition form factors. The
contributions of the TPE process to the $eN\rightarrow e\Delta(1232)\rightarrow eN\pi$  are calculated in a hadronic model with the inclusion of only the elastic nucleon intermediate states,  to estimate its effects on the multipoles $M^{(3/2)}_{1^+}, E^{(3/2)}_{1^+}, S^{(3/2)}_{1^+}$ at the $\Delta$ peak.  We find that TPE effects on $G^*_M$ is very small.  $G^*_E,$ and $ G^*_C$ are also little affected  at small $Q^2 $.  For  $G^*_E$, the TPE effects   reach about $3 - 8\%$ near $Q^2\sim 4$ GeV$^2$, depending on the model, MAID or SAID, used to emulate the data. For  $G^*_C$, the TPE effects   decrease rapidly with increasing $\epsilon$ while growing with increasing $Q^2$ to reach $\sim 6 - 15\%$ with $Q^2 \sim 4$ GeV$^2$ at $\epsilon = 0.2$. Sizeable TPE corrections to   $G^*_E$ and $G^*_C$
found here points to the need of including TPE effects in the multipole analysis  in the region of  high $Q^2$ and small $\epsilon$. The  TPE corrections to $R_{EM}$ and $R_{SM}$ obtained in our hadronic calculation are compared with those obtained in a partonic calculation for moderate momentum transfer of $2<Q^2<4$ GeV$^2$.
\end{abstract}
%\pacs{25.30.Bf, 12.20.Ds, 13.40.Gp}
%\textbf{Key words:} Two-Photon Exchange, $\Delta$
%\textbf{Key words:} Two-Photon Exchange, Form Factor
\maketitle
%%%%%%%%%%%%%%%%%%%%%%%%%%%%%%%%%%%%%%%%%%%%%%%%%%%%%%%%

The Jones-Scadron form factors,   magnetic dipole $G^*_M$, electric quardrupole $G^*_E$,
 and Coulomb quardrupole $G^*_C$, which describe electromagnetic transition between the first two lowest baryon states, nucleon and the $\Delta(1232)$ resonance, are of fundamental interest.  They are proportional to the three multipoles $M_{1^+}^{(3/2)}, E_{1^+}^{(3/2)},
 S_{1^+}^{(3/2)}$ at the resonance peak \cite{Pascal07}, which are all purely imaginary. Namely, on the resonance peak $W=M_\Delta$, one has
\begin{eqnarray}
&&G^*_M=N\, {\rm Im}\,M_{1^+}^{(3/2)}(Q^2,W=M_\Delta), \nonumber\\
&&G^*_E=-N\, {\rm Im}\,E_{1^+}^{(3/2)}(Q^2,W=M_\Delta), \nonumber\\
&&G^*_C=-(\frac{2M_\Delta}{q_\Delta})N\, {\rm Im}\,S_{1^+}^{(3/2)}(Q^2,W=M_\Delta),
 \label{ff-multipoles}
\end{eqnarray}
where $N=\frac{8}{e}(\frac{\pi k_\Delta M_\Delta \Gamma_\Delta}{3q_\Delta}\frac{Q_+}{Q_-})^{1/2}(\frac{M_N}{M_N+M_\Delta})$, with $e^2/4\pi\simeq 1/137$,  $Q_\pm\equiv [(M_\Delta\pm M_N)^2 + Q^2]^{1/2}$, $\Gamma_\Delta$ the $\Delta$ width, and $M_N$ and  $M_\Delta$ are the nucleon and $\Delta$ masses, respectively. $q_\Delta$ and $k_\Delta$ denote   the magnitude of the virtual photon and pion three momentum in the $\Delta$ rest frame   at the resonance position, respectively.

At sufficiently large four-momentum transfer squared $Q^2$,  perturbative QCD (pQCD)   predicts that only
helicity-conserving amplitudes
contribute  \cite{Brodsky81}, leading to $G^*_M, G^*_E, G^*_C$ scaling as $Q^{-4}, Q^{-4}$, and $Q^{-6}$, respectively.
It follows that
\begin{eqnarray}
\hspace{-0.5cm}
&& R_{EM} \equiv  ( E_{1^+}^{(3/2)}/M_{1^+}^{(3/2)})\mid_{W=M_\Delta} = -G^*_E/G^*_M \rightarrow 1, \nonumber \\
\hspace{-0.5cm}
&&R_{SM} \equiv  (S_{1^+}^{(3/2)}/M_{1^+}^{(3/2)})\mid_{W=M_\Delta} \nonumber\\
%&&\hspace{0.9cm}= -(\frac{Q_+Q_-}{4M_\Delta^2}) (G^*_C/G^*_M) \rightarrow const.
&&\hspace{0.9cm}= -(Q_+Q_-/4M_\Delta^2)(G^*_C/G^*_M) \rightarrow const.
 \label{Remsm}
\end{eqnarray}

In the nonpertubative regime with low
 $Q^2$, a symmetric SU(6) quark model would allow
the electromagnetic excitation of the $\Delta$ to proceed only
via $M1$ transition. However, the tensor component of the one-gluon exchange
interaction between quarks would induce a $D-$state in the $\Delta$, which leads to a deformed $\Delta$ and
the photon can excite a nucleon through electric $E2$ and Coulomb
$C2$ quardrupole transitions, resulting in  nonvanishing $E_{1^+}^{(3/2)}$ and
$S_{1^+}^{(3/2)}$ multipoles.   Experiments give,
near $Q^2 =0$, $R_{EM} = -(2.5\pm 0.5)\%$
\cite{Beck97}, a clear indication of $\Delta$ deformation. Below $Q^2 \le 6$ Gev$^2$,
$R_{EM}$ remains small and negative, while $R_{SM}$ continues to become more negative with
increasing $Q^2$, indicating that pQCD limit is nowhere in sight.
The intriguing difference in the behaviors of the $R_{EM}$ in the
 perturbative  and nonperturbative domains remains to
be understood.

The multipoles   are extracted from pion electroproduction
 experiments based on one-photon exchange (OPE) approximation. OPE
approximation has been widely used to analyze most of the electromagnetic nuclear reactions.
The validity of OPE approximation has recently been under heavy scrutiny  \cite{Carlson07, Arrington11,Yang13}.
It was prompted by the sustantial difference in the ratio of proton electric and magnetic form factors extracted from $ep$ elastic scattering via Rosenbluth technique
\cite{Arrington03,Qattan05} and   polarization transfer
  measurements  \cite{Jones00,Puckett10,Zhan11}, for $Q^2<6 $ GeV$^2$.  The two-photon exchange (TPE)
corrections as estimated by   hadronic and partonic calculations show that TPE effects can account for more than half of that discrepancy.

It is hence important  to determine how much TPE effects would affect the extraction of multipoles from pion   electroproduction.
Specifically we will be concerned with only the multipoles related to $N\Delta$ transition in this study, namely, how the extraction of
$E_{1^+}^{(3/2)}, M_{1^+}^{(3/2)}$, and $S_{1^+}^{(3/2)}$, or equivalently the transition form factors, would be affected in the presence of TPE. This question was addressed in \cite{Pascal06}, where  a partonic approach, with the
use of $N \Delta$ generalized parton distributions, was employed to estimate the TPE effects. For $2<Q^2<4$ GeV$^2$ at $\epsilon=0.2$, they found that the TPE corrections on $R_{EM}$ and $R_{SM}$, are small,
lying between  $-(0.2 - 0.6)\%$ level.  However, it is known that the partonic approach is applicable only for $Q^2$ large comparable to a typical
hadronic scale and becomes questionable  for $Q^2$, which in the current case, less than $\sim 2-3$ GeV$^2$. In these lower $Q^2$ region, hadronic approach as developed
in \cite{Blunden03} would be more reliable, which motivates this investigation.

\begin{figure}[htbp]
\center{\epsfxsize 1.7 truein\epsfbox{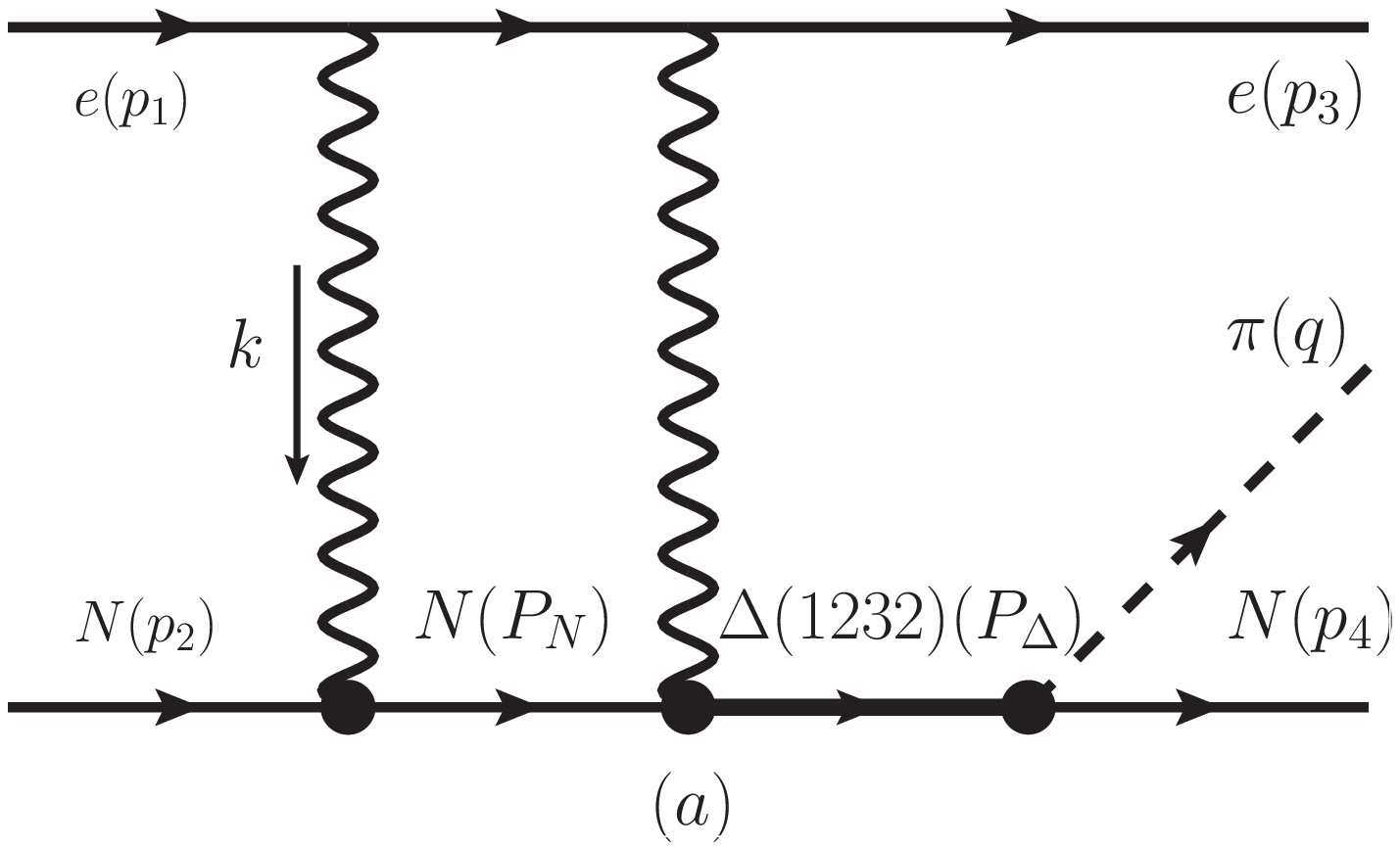}\epsfxsize 1.7 truein\epsfbox{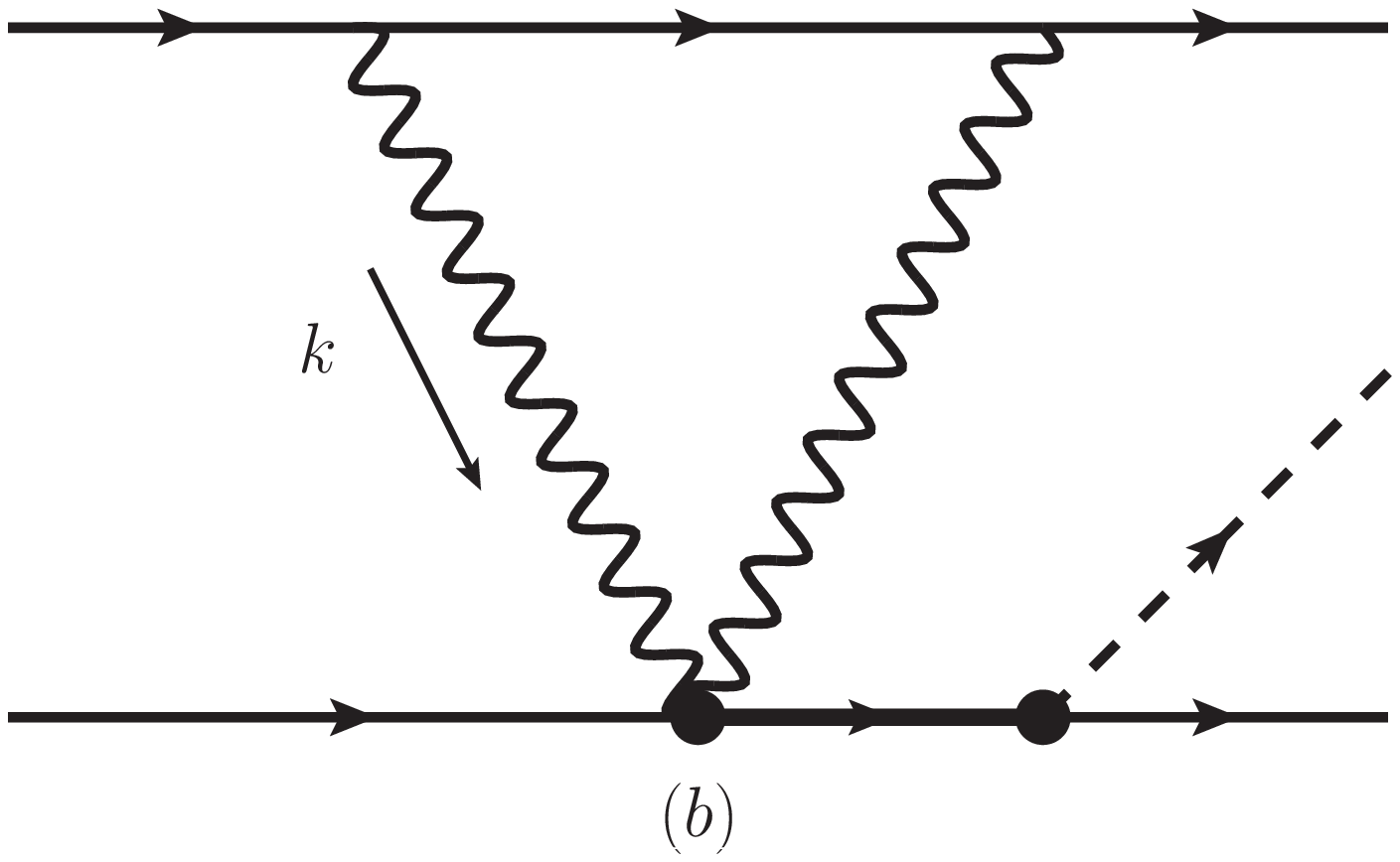}}
\caption{TPE box (a) and contact (b) diagrams  for $eN \rightarrow e\Delta \rightarrow eN\pi$. The cross-box diagram is not shown.}
\label{figure:ep-eppi0-TPE}
\end{figure}

In this work, we present results of a hadronic calculation of the TPE corrections, as depicted in Fig. \ref{figure:ep-eppi0-TPE}, where only the elastic $N$
intermediate states are considered,
to the process $eN\rightarrow eN\pi$ on the $\Delta$ peak. The intermediate nucleons are assumed to be on-mass-shell, which is justified in the study of TPE effects in $e p$ elastic
scatterings within hadronic approach in \cite{Blunden03}.

As in \cite{Zhou07,Zhou10,Zhou15}, we choose  Feynman gauge and neglect   electron mass $m_e$ in
the numerators to obtain the amplitude of box diagram in Fig. 1(a)  as
\begin{align}
&\mathcal{M}^{2\gamma, a}=-i\int\frac{d^4k}{(2\pi)^4}\overline{u}(p_3)\Gamma^{\gamma ee}_{\mu} S^{(e)}(p_1-k) \Gamma^{\gamma ee}_{\nu}u(p_1) \nonumber \\
&\times\frac{-i}{k^2+i\varepsilon}\frac{-i}{(P_\Delta-P_N)^2+i\varepsilon}\overline{u}(p_4) \Gamma^{\pi N \Delta,\alpha}(q)S^{(\Delta)}_{\alpha\beta}(P_\Delta)\nonumber \\
&\times \Gamma^{\mu \beta}_{\gamma N\rightarrow \Delta}(P_\Delta,P_\Delta-P_N)S^{(N)}(P_N)\Gamma^{\gamma NN,\nu}(k)u(p_2)
,\label{eq:diagram}
\end{align}
\noindent where $\Gamma^{\gamma ee}_\mu=-ie\gamma_{\mu}$, and $\Gamma^{\gamma NN}_\mu$=$ie\langle
P(p')|J^{em}_\mu|P(p)\rangle$, the  proton e.m. current matrix element.   $\Gamma^{\pi N \Delta,\alpha}(q)= (f_{\pi N \Delta}/m_\pi) q^{\alpha}T^\dagger$   denotes the $\pi N\Delta$ transition vertex function with $f^2_{\pi N \Delta}/4\pi=0.36$ and $T^\dagger$ the isospin $1/2\rightarrow 3/2$ transition operator.
$S^{(e,N,\Delta)}$ denote the propagators of electron, proton, and $\Delta$, respectively, as specified by the superscript. The forms of the $S^{(\Delta)}$ and the $\gamma N\rightarrow \Delta$ transition vertex function $\Gamma^{\mu \beta}_{\gamma N\rightarrow \Delta}$ can be found in \cite{Zhou15}. The realistic form factors are used for $\Gamma^{\gamma NN}_\mu$ and $\Gamma^{\mu \beta}_{\gamma N\rightarrow \Delta}$ as in \cite{Zhou15,Zhou10}.
Amplitude for the\emph{}
cross-box diagram  can be written down similarly. A contact term  $ M^{2\gamma, ct}$, as depicted in Fig. \ref{figure:ep-eppi0-TPE}b, is needed because of
the requirement of current conservation. Following the prescription  suggested in \cite{Kondra06}, we obtain

\begin{align}
&\mathcal{M}^{2\gamma, ct}=-i\int\frac{d^4k}{(2\pi)^4}\overline{u}(p_3)\Gamma^{\gamma ee}_{\mu} S^{(e)}(p_1-k) \Gamma^{\gamma ee}_{\nu}u(p_1) \nonumber \\
&\times\frac{-i}{k^2+i\varepsilon}\frac{-i}{(P_\Delta-P_N)^2+i\varepsilon}\overline{u}(p_4) \Gamma^{\pi N \Delta,\alpha}(q)S^{(\Delta)}_{\alpha\beta}(P_\Delta)\nonumber \\
&\times \Gamma^{\mu \nu \beta}_{\gamma \gamma N \Delta}(P_\Delta,p_2,k,p_4-p_2-k) u(p_2)
,\label{eq:diagram-ct}
\end{align}
with
\begin{align}
&\Gamma^{\mu \nu \beta}_{\gamma \gamma N \Delta}(P_\Delta,p_2,k,\overline{k})\nonumber \\
&=e\{(2p_2+k)^\nu\frac{F_1(k)}{(p_2+k)^2-M_N^2}  \Gamma^{\mu\beta}_{\gamma N\rightarrow \Delta}(P_\Delta,\overline{k}) \nonumber\\
&+(2p_2+\overline{k})^\mu\frac{F_1(\overline{k})}{(p_\Delta-k)^2-M_N^2}\Gamma_{\gamma N\rightarrow \Delta}^{\nu\beta} (P_\Delta,k)\},
\end{align}
where $\overline{k}=p_4-p_2-k$,  and $F_1$ is the Dirac form factor of the nucleon.
  The inclusion of   contact term of Eq. (\ref{eq:diagram-ct}) makes the full amplitude gauge invariant as discussed in  \cite{Kondra06}. We have also checked numerically that the full amplitude does not dependent on the gauge parameter. It is also essential  to ensure  the sum to be free of IR divergence. The packages
FEYNCALC \cite{Mertig91} and LoopTools \cite{Hahn99} are used to carry out the analytical
and numerical calculations, respectively.

Within the OPE
approximation,  the   fivefold $eN\rightarrow eN\pi$ differential cross section, with both unpolarized initial and final states,
can be  expressed as
$d^5\sigma^{1\gamma}/d\Omega_f dE_f d\Omega_\pi \equiv \Gamma d\sigma^{1\gamma}/d\Omega_\pi$,
 with $\Gamma$  the virtual photon flux factor and
\begin{equation}
 \frac{d\sigma^{1\gamma}}{d\Omega_\pi}= \{\sigma_0^{1\gamma}+\sqrt{2\epsilon(1+\epsilon)}\sigma_{LT}^{1\gamma}\cos\phi + \epsilon \sigma_{TT}^{1\gamma} \cos 2\phi \},
\label{unpolcross-section}
\end{equation}
where $\sigma_0^{1\gamma}=\sigma_T^{1\gamma}+\epsilon \sigma_L^{1\gamma}$, and  $\epsilon$   the transverse polarization of the virtual photon. The superscript $1\gamma$ is used
to emphasize that the quantities are defined within the OPE approximation scheme, a convention to be followed hereafter.
$E_f, \Omega_f$ denote the energy and
solid-angle of the scattered electron in the {\it lab} frame,
respectively, and $\phi$ is the tilt angle between the electron
scattering plane and the reaction plane, $d\Omega_\pi$ is the pion solid-angle
differential measured in the {\it c.m.} frame of the final pion
and nucleon.

The OPE differential cross sections
$\sigma^{1\gamma}_{T,L,LT,TT}$$'s$ are all functions of multipoles, which depend on $W$, $Q^2$ and pion polar angle $\theta_\pi$ in $\pi N$ c.m. frame, but $\epsilon$-independent.
The multipoles are determined in multipole analysis, e.g.,  MAID \cite{MAID07} or SAID \cite{SAID07}, by fitting the experimental data as,

\begin{eqnarray}
\frac{d\sigma^{ex}}{d\Omega_\pi} \simeq \frac{d\sigma^{1\gamma}}{d\Omega_\pi}  = C |\mathcal{M}^{1\gamma}(X^{1\gamma}_{1^+}, Z^{1\gamma}_{l^\pm})|^2
\label{multipoles-ope-a}
 \end{eqnarray}
where $ d\sigma^{ex}/d\Omega_\pi$
is measured experimentally. Here
$X^{1\gamma}_{1^+}=(E_{1^+}^{(3/2),1\gamma}, M_{1^+}^{(3/2),1\gamma}, S_{1^+}^{(3/2),1\gamma})$
denote the multipoles pertaining to the  $\Delta$ excitation channel of $(3/2, 3/2)$,
 $Z^{1\gamma}_{l^\pm}$ represents all other multipoles,
and $C$ is a kinematical factor.
%The superscript $M/S$ refers to whether they are obtained in MAID or SAID anaylsis.

 With  the TPE   effects included, the analysis of the experimental data should be performed by using,
\begin{eqnarray}
&&\frac{d\sigma^{ex}}{d\Omega_\pi} \simeq\frac{d\sigma^{1\gamma+2\gamma}}{d\Omega_\pi}
\nonumber \\
&&=C\{|\mathcal{M}^{1\gamma}(X^{1\gamma+2\gamma}_{1^+}, Z^{1\gamma+2\gamma}_{l^\pm})|^2\nonumber\\
&&+ 2Re[\mathcal{M}^{1\gamma *}(X^{1\gamma+2\gamma}_{1^+}, Z^{1\gamma+2\gamma}_{l^\pm})\mathcal{M}^{2\gamma}],
\label{multipoles-TPE}
 \end{eqnarray}
where the term $|\mathcal{M}^{2\gamma}|^2$ has been neglected.
 $(X^{1\gamma+2\gamma}_{1^+}, Z^{1\gamma+2\gamma}_{l^\pm})$    are the
multipoles determined from the OPE plus TPE approximation of Eq. (\ref{multipoles-TPE}), as referred to by the superscript
$1\gamma+2\gamma$, a notation to be followed hereafter.  Obviously, they
must deviate from
$(X^{1\gamma}_{1^+}, Z^{1\gamma}_{l^\pm})$
of Eq. (\ref{multipoles-ope-a})  based on OPE.
Eq. (\ref{unpolcross-section}) still holds for $d\sigma^{1\gamma+2\gamma}/d\Omega_\pi$
 but the cross sections $\sigma^{1\gamma+2\gamma}_{T,L, LT, TT}$'$s$ would
 become $\epsilon$-dependent \cite{Pascal07,Pascal06}.

In principle, one should try to determine the multipoles $X^{1\gamma+2\gamma}_{1^+}$ and $Z^{1\gamma+2\gamma}_{l^\pm}$ in the presence of TPE by fitting the data  with Eq. (\ref{multipoles-TPE}). The obtained values of the multipoles would represent the genuine multipoles as would be defined within the OPE approximation scheme, with TPE effects removed, from the data.

Extraction of  $X^{1\gamma+2\gamma}_{1^+}, Z^{1\gamma+2\gamma}_{l^\pm}$'s from data via Eq. (\ref{multipoles-TPE}) is beyond the scope of the present study.
To proceed, two approximations will be made. First, we  assume that only the
multipoles $X^{1\gamma+2\gamma}_{1^+}$$'s$ will be much affected in the presence of TPE depicted in Fig. \ref{figure:ep-eppi0-TPE}. This can be justified because
the final $\pi N$ pair there arised only from the decay of  $\Delta$ and would be in the state with $(J=3/2, I=3/2)$ only.    The multipoles
$Z^{1\gamma+2\gamma}_{l^\pm}$   will then be taken to be unchanged and
fixed, i.e., $Z^{1\gamma+2\gamma}_{l^\pm} = Z^{1\gamma}_{l^\pm}$   and Eq. (\ref{multipoles-TPE}) is reduced to depend only on the three multipoles of $X^{1\gamma+2\gamma}_{1^+} $$'s$. The Fermi-Watson theorem
requires that these three multipoles should all have the phase given by the $\pi N \,\, P_{33}$ phase shift, which is $\pi/2$ on the $\Delta$ peak. So   the three multipoles $X^{1\gamma+2\gamma}_{1^+}$'s will all become  purely imaginary in Eq. (\ref{multipoles-TPE}). Hereafter, $X_{1^+}$ will be taken to
denote the imaginary part of $X^{1\gamma+2\gamma}_{1^+}$ for brevity.

  Eq. (\ref{multipoles-TPE}) is then simplified to
\begin{eqnarray}
\frac{d\bar \sigma^{ex}}{d\Omega_\pi} &\equiv& \frac{d\sigma^{ex}}{d\Omega_\pi} -  2 C Re[\mathcal{M}^{1\gamma *}(X_{1^+})\mathcal{M}^{2\gamma}]\nonumber\\
&=&C|\mathcal{M}^{1\gamma}(X_{1^+})|^2,
\label{sigma-bar}
 \end{eqnarray}
 where a TPE-corrected cross section $d\bar \sigma^{ex}/d\Omega_\pi$ is introduced. Dependence on $Z_{l^\pm}$$'s$ in $\mathcal{M}^{1\gamma}$ in Eq. (\ref{sigma-bar}) is not shown for simplicity since they remain fixed. We like to emphasize here that the $\sigma^{ex}_{T, L, LT, TT}$$'s$, are in principle $\epsilon$-dependent. Only with precisely determined $d\sigma^{ex}/d\Omega_\pi$$'s$ and a complete theory for $\mathcal{M}^{2\gamma}$ would lead to $\epsilon$-independent $\bar\sigma^{ex}_{T,L, LT, TT}$$'s$.    $d\bar \sigma^{ex}/d\Omega_\pi$ is only then expressible in the form of $|\mathcal{M}^{1\gamma}|^2$.

To proceed, we   approximate the   data $d\sigma^{ex}/d\Omega_\pi$ with the use of one of the existing $eN\rightarrow eN\pi$ models,  MAID \cite{MAID07} or SAID \cite{SAID07}, to be denoted as
$\sigma^{MAID/SAID}_{T, L, LT, TT}$.
There is a caveat here with such an approximation.  All existing
models, like MAID and SAID, are all based on OPE approximation and the resulting cross sections $\sigma^{1\gamma}_{T, L, LT, TT}$$'s$ and multipoles
 would hence be $\epsilon$-independent. Approximating  $\epsilon$-dependent $\sigma^{ex}_{T, L, LT, TT}$$'s$   by   $\epsilon$-independent $\sigma^{MAID/SAID}_{T, L, LT, TT}$$'s$ would give rise to   $X_{1^+}$$'s$ determined from Eq. (\ref{sigma-bar}) to be  $\epsilon$-dependent.

Once
$d\sigma^{ex}/d\Omega_\pi$ is given, Eq. (\ref{sigma-bar}) then can be solved for $X^{1\gamma}_{1^+}$$'s$   by
iteration via,
\begin{eqnarray}
\frac{d\bar \sigma^{ex,i+1}}{d\Omega_\pi} &\equiv& \frac{d\sigma^{ex}}{d\Omega_\pi} -  2 C Re[\mathcal{M}^{1\gamma *}(X^{i}_{1^+})\mathcal{M}^{2\gamma}]\nonumber\\
&=&C|\mathcal{M}^{1\gamma}(X^{i+1}_{1^+})|^2.
\label{iteration}
\end{eqnarray}
We start  with values of multipoles given by MAID or SAID, i.e., $X^{0}_{1^+}=X_{1^+}$(MAID/SAID) in the first iteration $i=0$,
 depending on which model is employed to approximate $d\sigma^{ex}/d\Omega_\pi$ in  Eq. (\ref{sigma-bar}). It should be noted that both the
 {\it l.h.s.} and {\it r.h.s.} depend on $\theta_\pi$ and $\phi$.

Next, we have  to determine  the three multipoles $X^{i+1}_{1^+}$$'s$ from Eq. (\ref{iteration}) for fixed $Q^2$ and $\epsilon$ at the $i-$iteration. Upon first glance, one could
 in principle write down three  equations for each of the $\sigma_{0, LT, TT}$$'s$ and solve for the three variables $X_{1^+}$$'s$. These three equations  are all quadratic equations in $X_{1^+}$$'s$.  It turns out that   there are a few angles where no real solutions exist for this coupled algebraic equations. The solutions  show rapid  variations \textit{w.r.t.}
$\theta_\pi$ in the neighbourhood of these angles.
The reason can be traced to the approximation we make to replace $d\sigma^{ex}/d\Omega_\pi$ by ($d\sigma^{ex}/d\Omega_\pi$)(MAID/SAID) in  (\ref{iteration}).

We hence turn to least-square method. As reported in \cite{Zhou17},   results obtained with such minimization procedure show strong sensitivity to the angle-independent weights attached to each of the three cross sections $\sigma_{0, LT, TT}$$'s$.  We now understand that this sensitivity arises from the problem described in the last paragraph.
  Accordingly, we decide to follow the  fitting method adapted in MAID \cite{MAID07}. At the $i$-th iteraction, we minimize $\chi^2(Q^2,\epsilon)$ defined as
\begin{eqnarray}
\chi_i^2(Q^2,\epsilon) \equiv \sum_{\theta_{\pi},\phi } \Big(
\frac{\mathcal{N}_{i+1}}
{\delta d\sigma^{ex}(\theta_{\pi},\phi)} \Big)^2,
\end{eqnarray}
with
\begin{eqnarray}
\mathcal{N}_{i+1}&=&[d\bar\sigma^{ex,i+1}
  (\theta_{\pi},\phi)=C|\mathcal{M}^{1\gamma}(X^{i+1}_{1^+})|^2]\nonumber \\
&&-[\frac{d\sigma^{ex}}{d\Omega_\pi} -  2 C Re[\mathcal{M}^{1\gamma *}(X^{i}_{1^+})\mathcal{M}^{2\gamma}],
\label{chi2}
\end{eqnarray}
where $ d\sigma^{ex}/d\Omega_\pi=$($d\sigma^{ex}/d\Omega_\pi$)(MAID/SAID).   $X^{i}_{1^+}$'s are kept fixed while
$X^{i+1}_{1^+}$'s are varied in the minimization of $\chi^2_i$.
In Eq.(\ref{chi2}) $\delta d\sigma^{ex}(\theta_{\pi},\phi )$ is the total
 error of $d\sigma^{ex}(\theta_{\pi},\phi )$  which   also depends on $Q^2$ and $\epsilon$. In our  analysis,   the experimental errors at $Q^2=2.8$ GeV$^2$, $\epsilon=0.56$ and $Q^2=4$ GeV$^2$, $\epsilon=0.5$ provided in \cite{Frolov99},  are used. Either set of errors  give rise to
nearly identical results.
 We choose to use the ones
at $Q^2=2.8$ GeV$^2$, $\epsilon=0.56$  for all other values of $Q^2$ and $\epsilon$  considered.
\begin{figure}[htb]
\center{\epsfxsize 3.8 truein\epsfbox{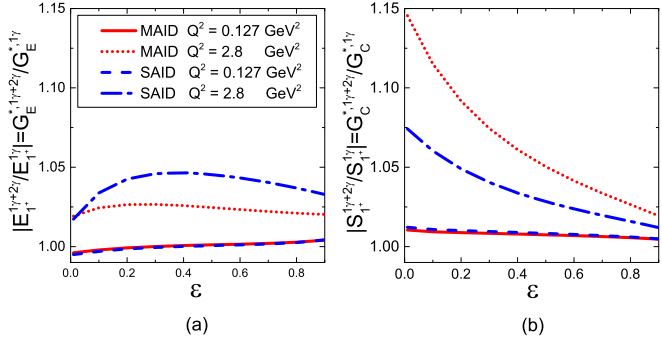}}
\caption{The TPE corrections to $E^{(3/2)}_{1^+}$ and $S^{(3/2)}_{1^+}$ \textit{vs.} $\epsilon$ at fixed $Q^2$. The labels MAID and SAID  are used to indicate that the results are obtained with using     either MAID or SAID to
emulate the experimental cross sections, respectively, as explained in the text. The solid and dotted  curves (red) refer to the results with MAID, the dashed  and dashed-dot  curves (blue) denote the results with SAID, respectively.  }
\label{Fig2}
\end{figure}

\begin{figure}[htb]
\center{\epsfxsize 3.8 truein\epsfbox{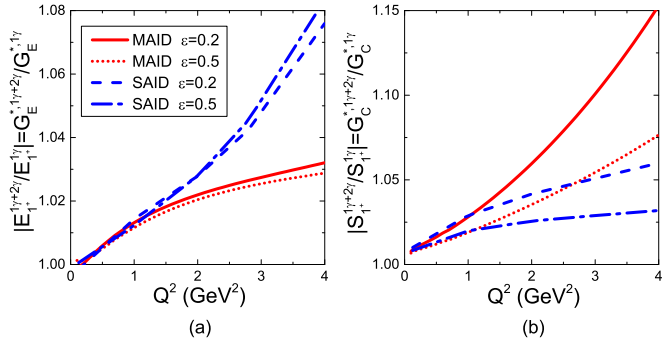}}
\caption{The TPE corrections to $E^{(3/2)}_{1^+}$ and $S^{(3/2)}_{1^+}$ \textit{vs.}  $Q^2$ at fixed $\epsilon$. Notation same as in Fig. 2.}
\label{Fig3}
\end{figure}

We  will show only the ratios $X^{1\gamma+2\gamma}_{1^+}/X^{1\gamma}_{1^+}\equiv G^{*, 1\gamma+2\gamma}/G^{*, 1\gamma}$ between the TPE-corrected, or the genuine OPE values  $X^{1\gamma+2\gamma}_{1^+} (\propto G^{*, 1\gamma+2\gamma})$ obtained here, and the
input OPE values $X^{1\gamma}_{1^+} (\propto G^{*, 1\gamma})$ given by the models (MAID, SAID) used to emulate the experimental
data. They will be labelled as MAID and SAID, respectively.  Results
 for $M^{3/2}_{1^+}$ will not be shown as the TPE effects on it are found to be very small
 with both models. We do not show results above $Q^2 > 4$ GeV$^2$ as the validity of hadronic approach adopted here might   be questionable in those high $Q^2$-region. The results, obtained with  MAID and SAID,  are presented for $0<\epsilon<0.9$ at $Q^2=0.127$ and 2.8 GeV$^2$, in Fig. 2, and for $0< Q^2 < 4$ GeV$^2$ with $\epsilon =$ 0.2 and 0.5,   in Fig.  3, respectively.  The results with MAID are
denoted by the solid   and dotted (red) curves, while the results with SAID are denoted by the dashed  and dashed-dot (blue) curves, respectively.

In Fig. 2, one sees that at small  $Q^2 = 0.127$ GeV$^2$, the TPE corrections to both  $E^{3/2}_{1^+} (G^*_E)$ and $S^{3/2}_{1^+} (G^*_C)$ are less than $1\%$ and stay flat for all values of {$\epsilon$}, irrespective of the model used. As $Q^2$ grows, TPE effects begin to increase and dependence on the model used develops.  For $E^{3/2}_{1^+} (G^*_E)$, the TPE corrections eventually reach about 3\% and 8\% at $4$ GeV$^2$ in the case of MAID and SAID, respectively, as seen in Fig. 3(a), with mild sensitivity {\it w.r.t.} $\epsilon$.
 The TPE corrections to $S^{3/2}_{1^+} (G^*_C)$ at $Q^2=2.8$ GeV$^2$, as depicted in Figs. 2(b) show considerable sensitivity not only to  model but also  $\epsilon$, decreasing from around 7.5\% and 15\% near $\epsilon=0$, for SAID and MAID, respectively, to only 2\% as $\epsilon$ approaches 0.9. Fig. 3(b) shows how TPE corrections for $S^{3/2}_{1^+} (G^*_C)$ grow with increasing $Q^2$ to reach about 15\% and 6\%, respectively, at $\epsilon = 0.2$ and  $Q^2=4$ GeV$^2$, for MAID and SAID. Sizeable TPE corrections to   $E^{3/2}_{1^+} (G^*_E)$ and $S^{3/2}_{1^+} (G^*_C)$
found here point to the need of including TPE effects in the multipole analysis of data in the region of  high $Q^2$ and small $\epsilon$.

It is straightforward to obtain the values for the TPE-corrected ratios $R^{1\gamma+2\gamma}_{EM,SM}$ from the results   presented in Fig. 3.  The difference $\delta R_{EM, SM}$ between  $R^{1\gamma+2\gamma}_{EM, SM}$
 and the model ratios $R^{1\gamma}_{EM, SM}$,  i.e., $\delta R_{EM} \equiv R^{1\gamma+2\gamma}_{EM}-R^{1\gamma}_{EM}$ and $\delta R_{SM}=R^{1\gamma+2\gamma}_{SM}-R^{1\gamma}_{SM}$, for $0 < Q^2< 4$ GeV$^2$ are shown in Fig. 4, where the solid (red) and dashed (blue) curves refer to the results obtained with MAID and SAID, respectively.    We first note that the TPE corrections $\delta R_{EM, SM}$ are almost equal with the two models except for $\delta R_{EM}$  when $Q^2 > 2$ GeV$^2$.   This is in contrast to Figs. 2 and 3  where   model dependence grows rapidly with increasing $Q^2$ after $Q^2 \sim 1$ GeV$^2$. For both $\epsilon =$ 0.2 and 0.5, $\delta R_{EM}$ is negligible for small $Q^2$ and becomes more negative toward $-0.1\%$ and $-0.2\%$ when $Q^2$ approaches $Q^2=4$ GeV$^2$, in the case of MAID and SAID, respectively.  The TPE effects for $\delta R_{SM}$ is considerably larger than for $\delta R_{EM}$. It also starts near zero for $Q^2 \sim 0$ but decreases rapidly to reach $\sim -1.4\%$ and $\sim -0.7\%$, for $\epsilon =$ 0.2 and 0.5, respectively. Magnitude-wise, they are comparable to the current experimental errors \cite{Ungaro06}.

The results of the partonic calculation of   \cite{Pascal06} for $\delta R_{EM/SM}$'s, denoted by    black triangles, are included in Fig. 4 for comparison. The regions of validity of the hadronic and partonic approaches are known to be
different except possible overlap in the range of $2<Q^2<4$ GeV$^2$. It is easily seen that, in this region, our results for $\delta R_{EM}$ at $\epsilon =$ 0.2 obtained with both models are
considerably smaller. However, for $\epsilon =$ 0.5, our results obtained with MAID almost coincide with those of \cite{Pascal06}, while results obtained with SAID are distinctly smaller than partonic results.   In the case
of $\delta R_{SM}$, our values are substantially more negative than the partonic results, for both $\epsilon = $ 0.2 and 0.5.
\begin{figure}[htbp]
\center{\epsfxsize 3.7 truein\epsfbox{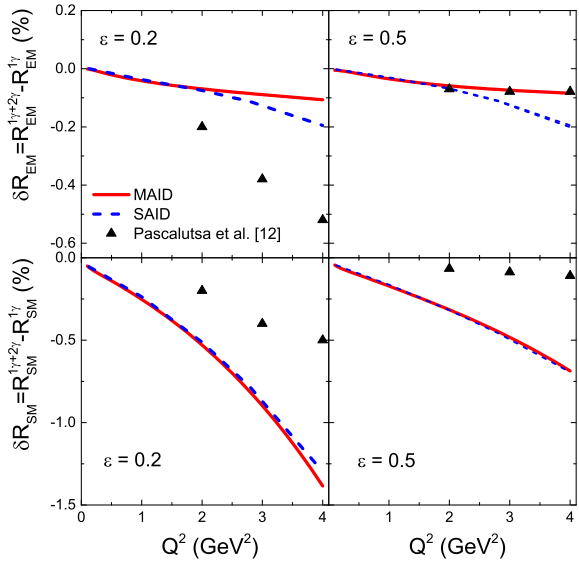}}
\caption{ The TPE corrections to the extracted $R_{EM}$ and $R_{SM}$ \textit{vs.} $Q^2$ at fixed $\epsilon$. The notation for   curves    same as in Fig. 2. The black triangles denote the results of the partonic calculation of \cite{Pascal06}.  }
\label{figure:REMRSM}
\end{figure}

The  differences between our results and those of \cite{Pascal06} for the $R_{EM, SM}$$'s$, as  shown in Fig. 4, can be dissected as follows.
We first point out that there are two more differences between the two calculations besides partonic \textit{vs}. hadronic approach. First,  only the $\Delta$ pole diagram
 is considered for $\mathcal{M}^{1\gamma}$ in \cite{Pascal06}, to evaluate the interference effects between OPE and TPE. In other words, the background contribution to  $\mathcal{M}^{1\gamma}$, which consists
 of Born terms in PV coupling and t-channel $(\rho, \omega)$ vector-meson exchanges \cite{Yang85}, are not included in the evaluation of $Re [\mathcal{M}^{1\gamma*}\mathcal{M}^{2\gamma}]$ in Eq. (\ref{sigma-bar}). In fact,
 it was found in \cite{KY99} that both the background terms and the pion cloud effects contribute significantly to $M^{(3/2)}_{1^+}$ and $E^{(3/2)}_{1^+}$ at $Q^2=0$. In addition, truncated multipole expansion (TME) is   employed in \cite{Pascal06} to estimate the values of $R^{1\gamma+2\gamma}_{EM, SM}$. It is known that  the use of TME and model fitting used here give rise to considerable difference in the extraction of $R^{1\gamma}_{EM, SM}$, a feature seen in \cite{Mertz01,Sparveris07}.

To summarize, we investigate the effects of two-photon exchange processes in $eN\rightarrow e\Delta(1232)\rightarrow eN\pi$ in   low $Q^2$ region,  in a hadronic approach. Only   the elastic nucleon intermediate states are included in the present study. We focus on the $\Delta$ peak to estimate their effects on the $\gamma^* N\Delta$ transition form factors. We emulate the experimental   pion electrproduction data with two existing phenomenological models, MAID and SAID. After subtracting out the interference of one-photon and two-photon exchanges from the data,
 the reminder is used to extract the "genuine" one-photon exchange multipoles $M^{(3/2)}_{1^+}, E^{(3/2)}_{1^+}, S^{(3/2)}_{1^+}$  at $W=M_\Delta$. This gives us the three $\gamma^* N\Delta$ form factors, $G^*_M$, $G^*_E$, and $G^*_C$, for $0<Q^2<4$ GeV$^2$.

We find that TPE effects on $G^*_M$ are very small. Both $G^*_E$ and $G^*_C$
 are also little affected  at small $Q^2< 0.5$ GeV$^2$. However, the TPE effects on $G^*_E$ and $G^*_C$ grow with $Q^2$ and the sensitivity {\it w.r.t.} $\epsilon$ and the data model used appears. For  $G^*_E$,
 the TPE effects   reach about 3\% and 8\% at $Q^2\sim 4$ GeV$^2$, depending on whether MAID and SAID is used to emulate the data, respectively, with mild dependence on $\epsilon$. For  $G^*_C$, the TPE effects  obtained with both MAID and SAID decrease rapidly with increasing $\epsilon$ while grow with increasing $Q^2$ and reach $\sim 15\%$ and  $\sim 6\%$ as $Q^2 \rightarrow 4$ GeV$^2$ at $\epsilon = 0.2$, respectively, for MAID and SAID. Sizeable TPE corrections to  $G^*_E$ and $G^*_C$
found here points to the need of including TPE effects in the multipole analysis of data in the region of  high $Q^2$ and small $\epsilon$.

 Our extracted
TPE corrections for $\delta R_{EM}\equiv R^{1\gamma+2\gamma}_{EM} - R^{1\gamma}_{EM}$ are very small at $\epsilon = 0.2$ and $0.5$,    for both   MAID and SAID models, up to $Q^2 \le 4.0$ GeV$^2$. This feature is similar  with results of the partonic calculation
of \cite{Pascal06}, except   our results are only about one third in magnitude given  in \cite{Pascal06} for $\epsilon = 0.2$.
However, our TPE corrections for $R_{SM}$, independent of the models used, are considerably larger in magnitude than the results of \cite{Pascal06}, reaching $\sim -1.4\%$ and $\sim -0.7\%$ for
$\epsilon = $ 0.2 and 0.5, respectively.

Besides hadronic $\it vs.$ partonic approach, the differences between our results and those of   \cite{Pascal06} for $\delta R_{EM/SM}$'s could be attributed to  two other simplifications used in \cite{Pascal06}. First,
in \cite{Pascal06}  only the $\Delta$ pole contribution is included in the OPE amplitude in the evaluation of the interference between OPE and TPE amplitudes. In addition, TME is invoked in the extraction of the rations $R_{EM/SM}$'s.

As the TPE effects on $G^*_E (\sim E^{(3/2)}_{1^+})$ and $G^*_C (\sim S^{(3/2)}_{1^+})$ found in this study are not small, more precision measurements
on $ep\rightarrow e\Delta(1232)\rightarrow ep\pi^0$  in the region of $2<Q^2<4$ GeV$^2$ will be very desirable.  It is important to have data taken
for the same $Q^2$ but at different values of $\epsilon$. The $\epsilon$-dependence in the resulting multipoles will be clear signature for the TPE effects.

 We have considered only the elastic nucleon intermediate states in the present study. Similar TPE effects arising from the inclusion of higher resonances like $\Delta$ in the intermediate states should be further
 pursued. TPE effects on the transition form factors of other higher resonances will also be an interesting question to explore.

%%%%%%%%%%%%%%%%%%%%%%%%%%%%%%%%%%%%%%%%%%%%%%%%%%%%%%%%%%

We  thank Dr. Lothar Tiator for helpful
communications, regarding MAID. We also thank Dr. T.-S.H. Lee for  careful reading of the manuscript and suggestions. This work is supported in part
by the National Natural Science Foundation of China under Grant No.
11375044, the Fundamental Research Fund for the Central Universities under Grant No. 2242014R30012 for H.Q.Z. and the National Science Council of the Republic of China (Taiwan) for S.N.Y. under grant No. NSC101-2112-M002-025.
H.Q.Z. would  like to gratefully acknowledge the support of
the National Center for Theoretical Science  of the National
Science Council of the Republic of China (Taiwan) for his visits in January, 2016 and February, 2017.
He also greatly appreciates the warm hospitality extended to him by the Physics Department of the
National Taiwan University during the visits.

\end{document}